\documentclass[aps,prl,reprint,superscriptaddress]{revtex4}
\usepackage{xcolor}
\usepackage{tikz}

\usepackage{amssymb,amsmath}
\usepackage{graphicx}
\usepackage{upgreek}
\usepackage[english]{babel}
\usepackage[T1]{fontenc}
\usepackage[utf8]{inputenc}
\usepackage{graphicx}
\usepackage{bm}
\def\pmb#1{\setbox0=\hbox{#1}
\kern-.025em\copy0\kern-\wd0 \kern-.05em\copy0\kern-\wd0
\kern-.025em\raise.0433em\box0}

\newcommand{\R}{\mathbb R}

\begin{document}
\title{Scattering problems in elastodynamics} 
\author{Andre Diatta}
\address{Aix$-$Marseille Universit\'e, CNRS, Ecole Centrale, Institut Fresnel, UMR CNRS 7249, 13013 Marseille, France}

\author{Muamer Kadic}
\address{Institute of Applied Physics, Institute of Nanotechnology, Karlsruhe Institute of Technology (KIT), 76128 Karlsruhe, Germany}
 
\author{Martin Wegener}
\address{Institute of Applied Physics, Institute of Nanotechnology, Karlsruhe Institute of Technology (KIT), 76128 Karlsruhe, Germany}

\author{Sebastien Guenneau}
\address{Aix$-$Marseille Universit\'e, CNRS, Ecole Centrale, Institut Fresnel, UMR CNRS 7249, 13013 Marseille, France}

\begin{abstract}
In electromagnetism, acoustics, and quantum mechanics, scattering problems can routinely be solved numerically by virtue of perfectly matched layers (PMLs) at simulation domain boundaries. Unfortunately, the same has not been possible for general elastodynamic wave problems in continuum mechanics. In this paper, we introduce a corresponding scattered-field formulation for the Navier equation. We derive PMLs based on complex-valued coordinate transformations leading to Cosserat elasticity-tensor distributions not obeying the minor symmetries. These layers are shown to work in two dimensions, for all polarizations, and all directions. By adaptative choice of the decay length, the deep subwavelength PMLs can be used all the way to the quasi-static regime. As demanding examples, we study the effectiveness of cylindrical elastodynamic cloaks of the Cosserat type and approximations thereof.
\end{abstract}
\maketitle


\section{Introduction}
Scattering of waves off objects is a central problem in physics \cite{Marcuvitz1994}. In recent years, it has gained additional interest in the context of cloaking \cite{Fleury2015,Cummer2016}, which aims at reducing or even eliminating scattering.  Amazingly, the deceptively simple case of continuum mechanics, which derives from Newton’s law and Hooke’s law, is among the most challenging cases. The challenge arises from the fact that waves in elastic media \cite{Graff1975} can have transverse, longitudinal, or mixed polarizations. Polarization conversion can occur, too. In sharp contrast, electromagnetic waves are usually transverse, acoustic waves are longitudinal, and quantum-mechanical matter waves are scalar. In addition, cloaking in elastodynamics requires elasticity tensors with broken minor symmetry that were not usually considered previously. To test new concepts and design future experiments based on complex spatially inhomogeneous and anisotropic elastic-material distributions, analytical solutions of the scattering problem are generally not available. Thus, obtaining reliable numerical solutions is crucial.

In computational electromagnetism, two significant advances during the past 35 years are vector-mixed finite elements (FEs) developed by Nédélec \cite{nedelec1980} and perfectly matched layers (PMLs) introduced in 1994 by Bérenger \cite{berenger1994} and by Chew and Weedon \cite{chew1994}. The latter have been extended to bianisotropic media by Teixeira and Chew \cite{teixeira1998}. Similar developments of PMLs occurred in elastodynamics \cite{chew1996,tromp2003} inspired by PMLs in electromagnetics.

Clayton and Engquist have paved the way for numerical investigation of scattering problems by explicitly considering an incident and scattered field \cite{Engquist1977,aubry1992} using numerical formulations employing 3D finite element and/or boundary element methods.
Refinements of boundary element methods include the  treatment of anisotropic unbounded media in elastodynamics, such as semi-infinite half-spaces, but this requires solving complex boundary integral equations \cite{Bonnet1995}.  However, in the tracks of \cite{Engquist1977,aubry1992}, a consistent infinitesimal finite-element cell method \cite{Wolf1996} which can be seen as a finite element based boundary element method, allows using one row of elements to model infinite domains, with an asymptotic treatment in the radial direction that naturally fulfills the outgoing waves' radiation conditions. In this way, one can handle complex heterogeneous anisotropic obstacles in unbounded elastic media.

Another type of problem in computational physics appeared in the past twenty years with the rapidly growing field of photonic and phononic crystals \cite{pcbook2,milton2000,pcbook1,khelif1,Maldovan2013,Zhu,Ertekin2014,dasilva2016}, and of course metamaterials \cite{kadic1,kadic2,craster}.  The computation of band diagrams requires the application of Bloch-Floquet theory, which is well developed in condensed matter physics, to electromagnetic and acoustic waves in periodic media. In fact, scattering problems have been studied for decades, mainly in the context of electromagnetism or acoustics. In these fields, the decomposition of incident waves and scattered waves was comparably straightforward because the waves are either purely transverse or purely longitudinal in polarization. 

In this paper, we propose a rigorous and easily implementable path to solve problems of diffraction and scattering in the context of elastodynamics using a Finite-Element Method (FEM). We first look at so-called adaptative perfectly matched layers that can efficiently attenuate elastodynamic waves within a deeply subwavelength region, without any reflection for all polarizations and incidence. We then move on to a general way to implement the scattering problem in elastodynamics in homogeneous linear non-dispersive media and how to compute the scattered field from an arbitrary object having continuity, stress-free or clamped boundary conditions at its surface. This combination brings us into a position to study the scattering of elastodynamic cloaks deduced from a geometric transform, leading to heterogeneous media of the Cosserat type. Although non-dispersive, such Cosserat cloaks are amongst the most complex cases to solve in terms of diffraction and scattering, within the framework of linear elastodynamics. Indeed, such cloaks are described by inhomogeneous and fully anisotropic non-symmetric rank-4 elasticity tensors (and heterogeneous isotropic densities) as required for both polarization conversion and coupling. 
(Note that alternative routes to elastodynamic cloaking exist that preserve the symmetries of the elasticity tensor, such as using Willis' equations \cite{milton2006} or transformed pre-stressed solids \cite{norris2012}.) We finally investigate the numerical implementation of the Cosserat cloak.

\section{Adaptative Perfectly Matched Layers}

In the absence of a source, we usually write the Navier equations for the total displacement field ${\bf u}$ for time-harmonic excitation as
\begin{eqnarray}\label{nav1}
\nabla\cdot \left[ {\mathbb C} :\nabla  {\bf  u}
 \right]  + \rho\omega^2{\bf  u}={\bf 0}
\end{eqnarray}
where $ {\mathbb C}$ is the (symmetric) elasticity tensor, $\rho$ the mass density and $\omega$ the angular frequency of the wave.



\begin{figure}[h!]
\centering \includegraphics[scale=0.76]{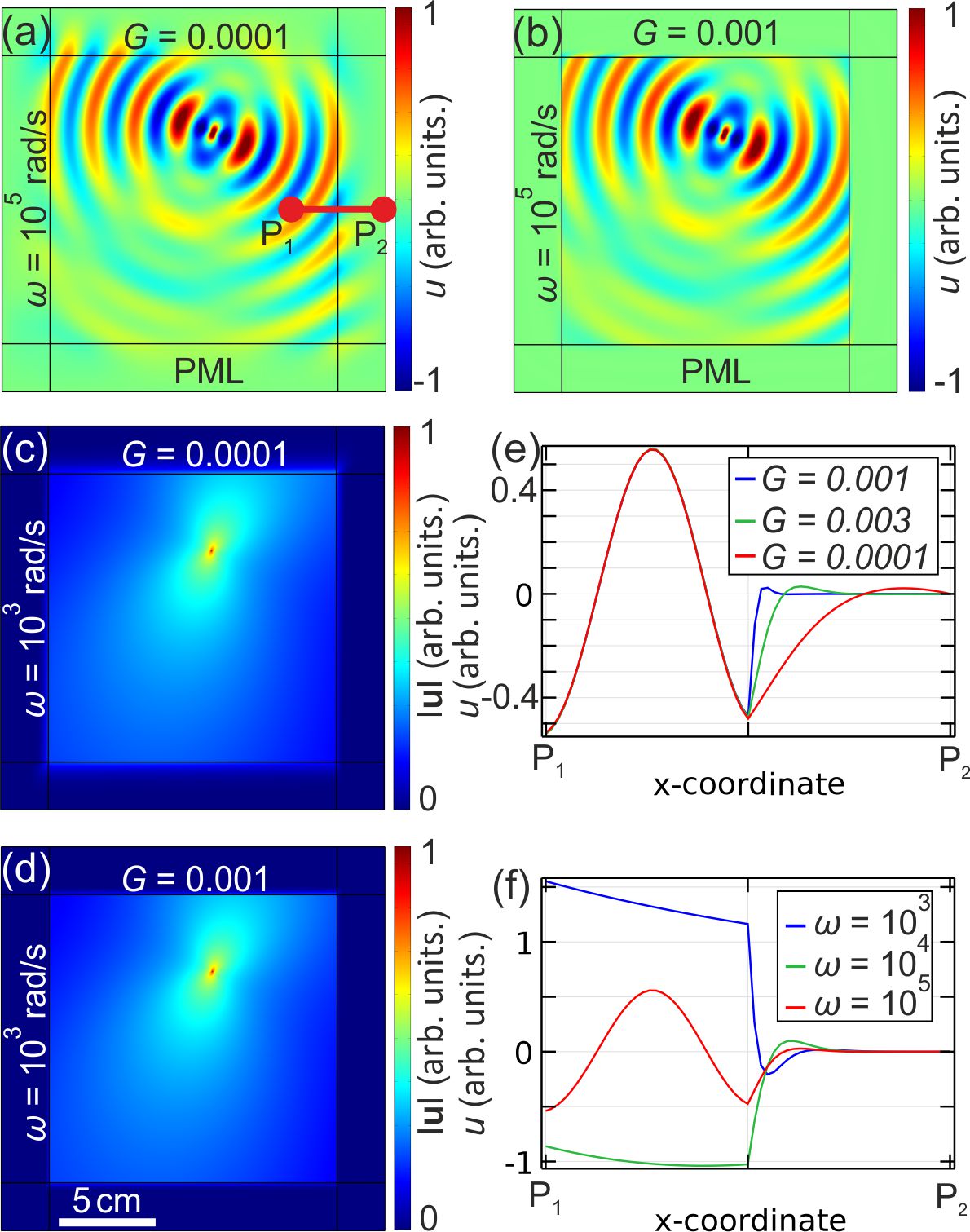}
\caption{Performance of adaptative PMLs: (a) Real part of the $u$-component of the displacement field ${\bf u} = (u, v)$ for a point source oscillating at an angular frequency $\omega=10^{3}$ rad.s$^{-1}$ and PMLs' scaling factor $G=10^{-4}$. (b) Same as (a), but for $G=10^{-3}$. (c), (d) Same as (a) and (b), respectively, for modulus of the displacement field.
(e), (f) Profile of $v$ along segment $[P_1,P_2]$ (see panel (a)) demonstrating the performance of PMLs. In (e) the scaling factor $G$ is varied for fixed $\omega=10^{3}$rad.s$^{-1}$. In (f) the angular frequency is varied and $G=0.003$ is fixed.}
\label{fig1}
\end{figure}

Before we can analyze the scattering of an arbitrary object, we need to be able to model scattering problems in unbounded domains. Owing to their ability to strongly absorb incoming waves in a reflectionless manner, PMLs help to model, within bounded domains, problems with open boundaries. Implementing general PMLs in elastodynamics is tricky. This problem becomes even more challenging in the quasi-static limit, where the wavelength tends to infinity and the entire radiated field is in the near field. Thus, one needs sufficiently large PMLs in order to enforce the decay of the elastic displacement-vector field down to zero on the outer boundary of the computational domain. Here, we propose a type of adaptative elastic PMLs, which are well suited for dealing with cases ranging from the quasi-static limit to high-frequency settings, as illustrated in Fig. \ref{fig1}. Our approach is inspired by earlier work in electromagnetism \cite{demesy2007} and is obtained from transformational techniques \cite{diatta2014} applied to the Navier equations  (\ref{nav1}), using  
 the transformation
\begin{eqnarray}\label{transfo:pml}
  (x',y',z')  =\Big(
x_0+\int_{x_0}^xs_1(\xi)d\xi,  
y_0+\int_{y_0}^ys_2(\xi)d\xi, 
z_0+\int_{z_0}^zs_3(\xi)d\xi
\Big).\nonumber
 \end{eqnarray}  
The $(x',y',z')$ are the working complex coordinates and the stretches $ s_{1}(\xi) , s_{2}(\xi) , s_{3}(\xi) $ are either equal to 1 or to 
$(\xi/L) \, \, (1-i) G, $   
depending on the direction along which one would like to absorb the wave, in a given region, where $i$ is the complex number with $i^2=-1.$ 
 The width of the PML region $L$ is a geometrical parameter that is automatically extracted for each region, whereas, the dimensionless PMLs scaling factor 
$G$, possibly incompassing a frequency dependence,  can be modified at will in order to achieve the needed PMLs efficiency (see Fig. \ref{fig1}).
This transformation is mapped onto the coefficients $C^{\rm pml}_{ijkl}$ of the elasticity tensor $\mathbb{C}^{\rm pml}$ and the mass-density tensor $\rho^{\rm pml}$ in the PML region \cite{diatta2014}. These coefficients have been implemented by us in the PDE (Partial Differential Equation) interface of the commercially available software COMSOL Multiphysics which is used for all computations in this paper.

For a bounded 2D isotropic homogeneous elastic medium with Lam\'e coefficients $\lambda,$ $\mu,$ the $8$ nonvanishing elastic coefficients in the PML region read: 
\begin{eqnarray}
C^{\rm pml}_{1111}&=&\frac{s_2(y)}{s_1(x)}(\lambda+2\mu), ~ 
C^{\rm pml}_{1122}=C^{\rm pml}_{2211}=\lambda,~
\nonumber\\   
C^{\rm pml}_{1212}&=&\frac{s_2(y)}{s_1(x)}\mu,~
 C^{\rm pml}_{1221}=  C^{\rm pml}_{2112}=\mu,~ 
\nonumber\\ 
 C^{\rm pml}_{2121}&=&\frac{s_1(x)}{s_2(y)}\mu,~  C^{\rm pml}_{2222}=\frac{s_1(x)}{s_2(y)}(\lambda+2\mu), \nonumber
\end{eqnarray}
and the density is as follows
\begin{equation}
\rho^{\rm pml}={s_1(x)}{s_2(y)}\; \rho \; .\nonumber
\end{equation} 

\section{Scattered field formulation}

Next, we consider the problem of an elastic wave impinging onto an obstacle $\Omega\subset\R^n$, $n=2,3$, resulting in scattering off of that object. We write the total displacement-vector field as the sum of the incident and the scattered fields. Solving the Navier equation (1) for the incident field
\begin{eqnarray}
\nabla\cdot \left[  {\mathbb C} :\nabla {\bf  u}^{\rm inc}
 \right]  + \rho\omega^2 {\bf  u}^{\rm inc}={\bf 0} \; ,
\label{navinc}
\end{eqnarray}
is straightforward for plane waves, cylindrical waves, and spherical waves. We shall restrict ourselves to incident plane waves in this paper.

\subsection{Arbitrary scatterer with continuity boundary conditions}
Following \cite{demesy2007}, let  us now consider a (possibly piecewise constant heterogeneous) obstacle $\Omega$ described
by an elasticity tensor $ {\mathbb C}_0$ and a density $\rho_0$.
Let $\tilde {\mathbb C}$ be a 4th-order elasticity tensor defined in $\R^n$ as 
$\tilde {\mathbb C}:= {\mathbb C}{\bf 1}_{\R^n\setminus \overline{\Omega}}+ {\mathbb C}_{0}{\bf 1}_{\Omega}$
 where ${\bf 1}_{\Omega},$ and  ${\bf 1}_{\R^n\setminus\overline{\Omega}},$ are the indicator functions with
support in $\Omega$ and $\R^n\setminus \overline{\Omega}$, respectively, where $\bar\Omega$ denotes the closure of $\Omega$.
Similarly, we consider the scalar density
$\tilde\rho:=\rho{\bf 1}_{\R^n\setminus \overline{\Omega}}+\rho_{0}{\bf 1}_{\Omega}$.

\noindent The total displacement field $\tilde{\mathbf u}$ is solution of 
\begin{eqnarray}\label{utotal1}
\nabla\cdot \left[\tilde  {\mathbb C} :\nabla  \tilde{\bf  u}
 \right]  + \tilde\rho\omega^2\tilde{\bf  u}={\bf 0} 
\end{eqnarray}
with $\tilde{\mathbf u}= {\bf  u}^{\rm inc}+{\bf u}^{\rm sc}$, where ${\bf  u}^{\rm inc}$ is the incident and ${\bf u}^{\rm sc}$ the scattered field (the latter satisfying outgoing wave conditions). Equation (\ref{utotal1}) becomes
\begin{eqnarray}
\nabla\cdot \left[\tilde  {\mathbb C} :\nabla ( {\bf  u}^{\rm inc}+{\bf u}^{\rm sc})
 \right]  + \tilde\rho\omega^2\Big({\bf  u}^{\rm inc}+{\bf u}^{\rm sc}\Big)={\bf 0} .
\label{navtot}
\end{eqnarray}
As the incident field ${\bf  u}^{\rm inc}$ satisfies (\ref{navinc}), (\ref{navtot}) leads to
\begin{eqnarray}
\nabla\cdot \left[ \tilde {\mathbb C} :\nabla  {\bf u}^{\rm sc}
 \right]  + \tilde\rho\omega^2 {\bf u}^{\rm sc}= {\bf F}
\label{navscat}
\end{eqnarray}
where the source term ${\bf F}$ is defined as
\begin{eqnarray}\label{diffractingmember}
{\bf F}=
 \nabla\cdot \left[ \Big( {\mathbb C} -\tilde {\mathbb C}\Big) :\nabla  {\bf  u}^{\rm inc}
 \right]  + \Big(\rho-\tilde\rho\Big)\omega^2~{\bf u}^{\rm inc} \; .
\end{eqnarray}
Importantly, ${\bf F}$ has a compact support in $\Omega$, which means that the scattered field problem amounts to solving the Navier equations with sources inside the obstacle $\Omega$. In other words, $\Omega$ acts as a virtual antenna.
Of course, ${\bf F}$ is known, since ${\bf  u}^{\rm inc}$ can be computed analytically.
Last, but not least, we note that the solution of (\ref{navscat}) allows for a completely
reflectionless implementation of elastic PMLs, such as those used in \cite{diatta2014, diatta2016}. In fact the fictitious source is now inside the computational domain and can thus be computed using FEM with PMLs.


\subsection{Stress-free obstacle}

Let us detail how this scattered-field formulation can be implemented in the two important cases of stress-free and clamped boundary conditions. We first consider a stress-free obstacle occupying a domain $\Omega$ of $\R^n.$ The scattering field problem is now defined in $\R^n\setminus\bar\Omega$ and we obtain
\begin{eqnarray}
\nabla\cdot \left[  {\mathbb C} :\nabla  {\bf u}^{\rm sc}
 \right]  + \rho\omega^2 {\bf u}^{\rm sc}= {\bf 0}
\label{navscatperforated}
\end{eqnarray}
with the source term now defined on the boundary $\partial\Omega$ of $\Omega$ as
\begin{eqnarray}\label{scat-stress-free}
 \left[ {\mathbb C} :\nabla  {\bf  u}^{\rm sc}
 \right] \cdot {\bf n}\mid_{\partial\Omega} =  -\left[ {\mathbb C} :\nabla  {\bf  u}^{\rm inc}
 \right] \cdot {\bf n}\mid_{\partial\Omega} \; .
\end{eqnarray}

The case of a clamped obstacle is straightforwardly deduced by replacing (\ref{scat-stress-free}) with
\begin{eqnarray}\label{scat-clamped}
{\bf  u}^{\rm sc} \mid_{\partial\Omega}=  -{\bf  u}^{\rm inc}\mid_{\partial\Omega}\,,
\end{eqnarray}
in which case all of the above equations simplify.

\begin{figure}[h!]
\centering \includegraphics[scale=0.76]{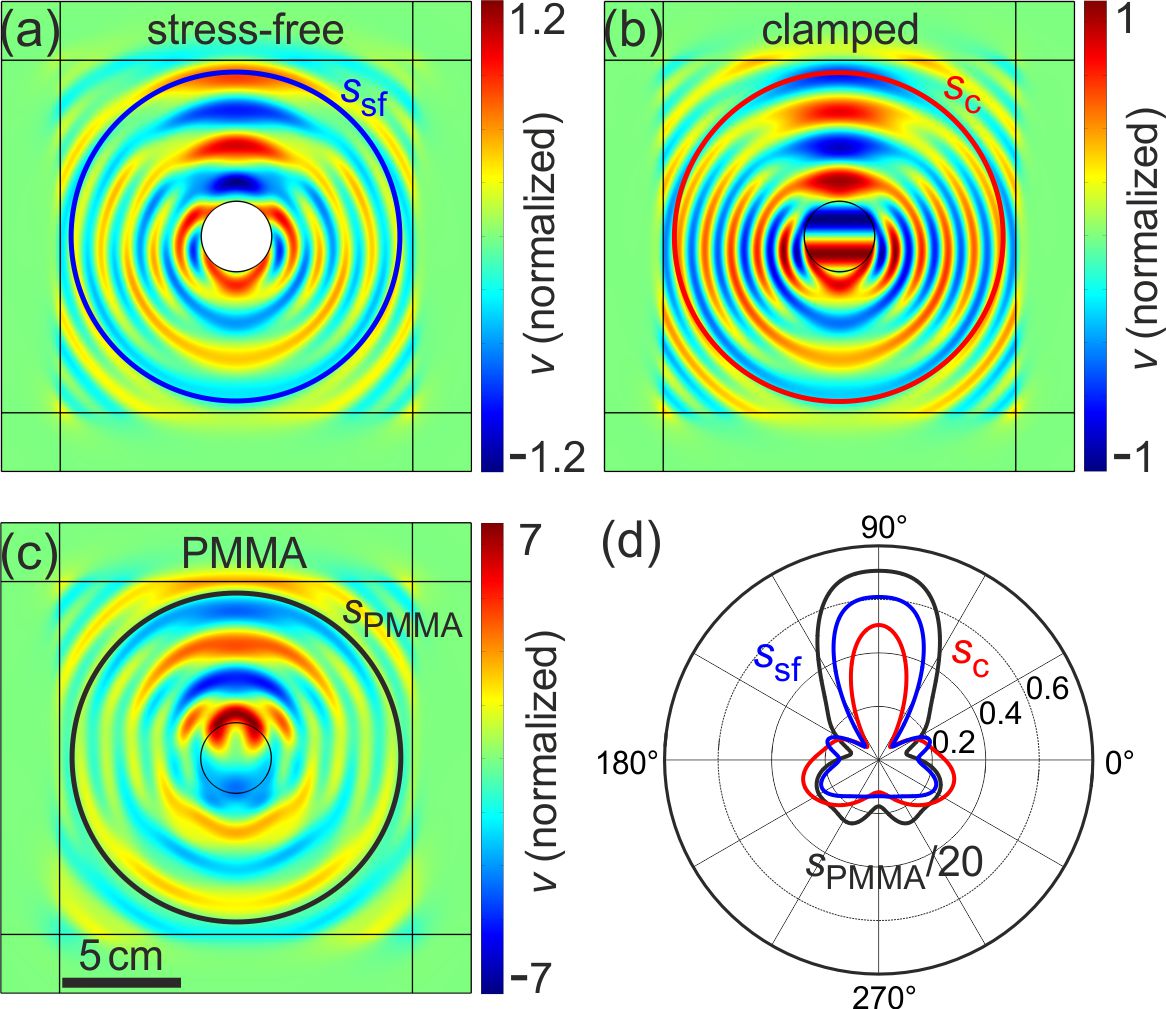}
\caption{Scattering for an incoming compression plane wave emanating from the top at frequency $2\times 10^5$ rad.$\rm s^{-1}$, for  (a) stress-free,  (b) clamped and  (c) solid obstacles. The solid obstacle in (c) has a density of $1000$ kg/m$^3$, a first Lam\'e coefficient $\lambda=4.3$ GPa and a shear modulus  $\mu=1$ GPa. The ambient material surrounding the obstacle has a first Lam\'e coefficient $\lambda=0.5$ GPa,  a shear modulus   $\mu=0.1$ GPa  and a mass density $200$ kg/m$^3$.
(d) Polar plot of ${\mid{\bf u}\mid}^2=u\bar{u}+v\bar{v}$, with $\bar{z}$ the complex conjugate of $z$, computed on the circle with radius $r_3$, which is depicted in panels (a)-(c).
}
\label{fig2}
\end{figure}

In Fig.\,\ref{fig2}, we show two-dimensional examples of the real parts of scattered fields for a stress-free (a), a clamped (b), and a solid (in polymethylmethacrylate or PMMA, for short)  (c) obstacles subject to an incident  compressional plane wave. The polar plot of $|{\bf u}|^2=u\bar{u}+v\bar{v}$ in panel (d) clearly shows that scattering by solid PMMA is the most pronounced case, followed by the clamped case and then the stress-free case.

\subsection{Application to ideal Cosserat and approximated cloaks}

As a more demanding example, we consider the case where the scattering obstacle $\Omega$ is surrounded by an invisibility cloak $\Omega_f$. 
The scattered field formulation (\ref{navscat})-(\ref{diffractingmember}) still holds within the obstacle $\Omega$, but one needs to consider the cloak $\Omega_f$  as a scattering obstacle as well, for which (\ref{navscat}) also applies. However, now the new tensor  $\tilde{\tilde{\mathbb C}}$  and mass density $\tilde{\tilde{\rho}}$  must take into account the cloak,
\begin{eqnarray}
\nabla\cdot \left[ \tilde{\tilde{\mathbb C}} :\nabla  {\bf u}^{\rm sc}
 \right]  + \tilde{\tilde\rho}\omega^2 {\bf u}^{\rm sc}= {\bf F}_f
\label{navscat1}
\end{eqnarray}
where the source term ${\bf F}_f$ is defined as
\begin{eqnarray}
{\bf F}_f=
 \nabla\cdot \left[ \Big({\mathbb C} - \tilde{\tilde{\mathbb C}}\Big) :\nabla  {\bf  u}^{\rm inc}
 \right]  + \Big(\rho-\tilde{\tilde\rho}\Big)\omega^2~{\bf  u}^{\rm inc} \; .
\end{eqnarray}
Precisely, $\tilde{\tilde{\mathbb C}}$ is a 4th-order asymmetric elasticity tensor defined in $\R^n$ as 
$\tilde{\tilde{\mathbb C}}:={\mathbb C}{\bf 1}_{\R^n\setminus \overline{\Omega\cup\Omega_f}}+{\mathbb C}_{f}{\bf 1}_{\Omega_f}$,
where ${\bf 1}_{\Omega_f},$ and  ${\bf 1}_{\R^n\setminus\overline{\Omega\cup\Omega_f}},$ are the indicator functions with
support in $\Omega_f$ and $\R^n\setminus \overline{\Omega\cup\Omega_f}$, respectively.
Similarly, we consider the scalar density
$\tilde{\tilde\rho}=\rho{\bf 1}_{\R^n\setminus \overline{\Omega\cup\Omega_f}}+\rho_{f}{\bf 1}_{\Omega_f}$.

\noindent In Cartesian coordinates, the spatially varying transformed elasticity tensor ${\mathbb C}_{f}$ is defined as 

\begin{eqnarray}{  C}^{f}_{ijkl}=\Big({\frac{\partial f_1}{\partial x_1} \frac{\partial f_2}{\partial x_2}-\frac{\partial f_1}{\partial x_2} \frac{\partial f_2}{\partial x_1}}\Big)^{-1}\frac{\partial f_i}{\partial x_p} \frac{\partial f_k}{\partial x_q}C_{pjql},\nonumber
\end{eqnarray}
 where $i,j,k,l,p,q=1,2$ (with an implicit summation on repeated subscripts) and $f$ is the transformation defining the cloak. In the case where a scattering stress-free obstacle $\Omega$ (object to cloak) shares a boundary $\partial\Omega$ with the cloak (inner boundary of the cloak), the source term (\ref{scat-stress-free}) is defined on $\partial\Omega:$
\begin{eqnarray}
 \left[ \tilde{\tilde{{\mathbb C}}} :\nabla  {\bf  u}^{\rm sc}
 \right] \cdot {\bf n}\mid_{\partial\Omega} =  -\left[ \tilde{\tilde{{\mathbb C}}} :\nabla  {\bf  u}^{\rm inc}
 \right] \cdot {\bf n}\mid_{\partial\Omega} \; .
\label{bc1}
\end{eqnarray}

We further consider the example of a linear radial transformation 
\begin{eqnarray}
f(r,\theta)= (r',\theta')=\Big( r_1+\frac{(r_2-r_1)}{r_2}r\; , \; \theta\Big)
\nonumber
\end{eqnarray}
 in polar coordinates, first proposed as a design tool for a cylindrical electromagnetic cloak in \cite{Schurig2006}.
Here, $r_1$ and $r_2$ are the inner and outer radius of the cloak, respectively. This transformation maps a disc of radius $r_2$ to an annulus of inner and outer radii $r_1$ and $r_2$ respectively, the center $(0,0)$ to a circle of a radius $r_1$ and fixes point-wise the circle of radius $r_2$. This transformation leads to the Cosserat  elasticity-tensor distribution  for the cloak, which explicitly reads:
\begin{eqnarray}C_{r'r'r'r'}'&=&a(\lambda+2\mu),  ~~
C_{r'r'\theta\theta}'=C_{\theta\theta r'r'}'=\lambda,  ~~
 C_{r'\theta r'\theta}'= a\mu,\nonumber
\\
 C_{r'\theta\theta r'}' &=&C_{\theta r'r'\theta}' =\mu, ~~
 C_{\theta r'\theta r'}'=\ \frac{\mu}{a}, ~~
C_{\theta\theta\theta\theta}'=\frac{1}{a}(\lambda+2\mu),
\nonumber
\end{eqnarray}
 where  $a=\frac{r'-r_1}{r'}$, and to the mass density 
\begin{eqnarray}\rho'= \Big(\frac{r_2}{r_2-r_1}\Big)^2a\rho \;.
\nonumber
\end{eqnarray} Here we have simply  recast $C_{ijkl}^f$ as  $C_{ijkl}'$, for $i,j=r',\theta,$ 
 \cite{Brun2009}.

Let us now apply our scattered-field formalism combined with the PML to numerical studies of elastodynamic cloaking. First, we consider the ideal Cosserat cloak. By construction, it should exhibit ideal cloaking. Second, we consider a symmetrized version of the elasticity tensor. 
The motivation for the latter two is to reduce the complexity and the requirements of the cloak, if possible. Indeed, homogenization theory shows it is only possible to achieve symmetric homogenized tensors with concentric layered media \cite{jikov}.

Figures 3(a) and 3(b) show the scattered field and the total field, respectively, for an obstacle such as in Fig.\,2(c) and for an incident compression wave at an angular frequency of $2\times 10^5$ rad.$\rm s^{-1}$. The Cosserat cloak in Fig. 3 (c) shows a scattered field that is confined to the cloak and a recovery of the incident plane wave in Fig. 3(d).  This is an  explicit numerical demonstration for elastodynamic cloaking. Cloaking of a similar quality has been obtained for many other frequencies  for both incident compression and incident shear waves (not depicted). However, is the complicated non-symmetric Cosserat-material distribution really necessary to obtain good elastic cloaking? To investigate this question, we consider as the second example a symmetrized, i.e., non-Cosserat elasticity-tensor distribution. Note that this symmetrization is not unique, but any chosen symmetrization must be physical in that all eigenvalues of the resulting elasticity tensor need to be real or complex valued with positive imaginary parts, at least for passive media. For a possible symmetrization, the coefficients, again in polar coordinates, are given by
\begin{eqnarray}
C_{r'r'r'r'}^{\rm sym}&=&a(\lambda+2\mu), \nonumber\\
C_{r'r'\theta\theta}^{\rm sym}&=&C_{\theta\theta r'r'}^{\rm sym}=\lambda,  \nonumber\\
 C_{r'\theta r'\theta}^{\rm sym}&=& C_{r'\theta\theta r'}^{\rm sym} =C_{\theta r'r'\theta}^{\rm sym} =C_{\theta r'\theta r'}^{\rm sym} =\beta\mu, 
 \nonumber\\
C_{\theta\theta\theta\theta}^{\rm sym}&=&\frac{1}{a}(\lambda+2\mu),\nonumber
\end{eqnarray}
 with the dimensionless parameter $\beta=\frac{1}{4}(2+a+\frac{1}{a}).$ 
 The abbreviation $a$ has been given above. 

The corresponding numerical results shown in Figs.\,3(e) and 3(f) do show a large qualitative improvement with respect to the obstacle case shown in Figs.\,3(a) and 3(b). However, we also find superimposed wiggles compared to the ideal case in Figs.\,3(c) and 3(d). These wiggles originate from shorter-wavelength shear-like partial waves, which are generated by the approximative cloak and which interfere with the incident compression-like wave. The cloaking quality can be quantified by applying measures such as those introduced in \cite{buckmann2015}. Upon applying such hard quality measures, due to the presence of the wiggles, we find hardly any improvement of the approximative cloak. This statement holds true for angular frequencies $\omega$ ranging from $10^3$ to $3\times 10^5$ rad.s$^{-1}$ (not depicted). Other symmetrizations of the elasticity tensor lead to comparable results.  This suggests that in order to achieve cloaking with a symmetric elasticity tensor, the density needs to be anisotropic.

\begin{figure}[h!]
\centering \includegraphics[scale=0.8]{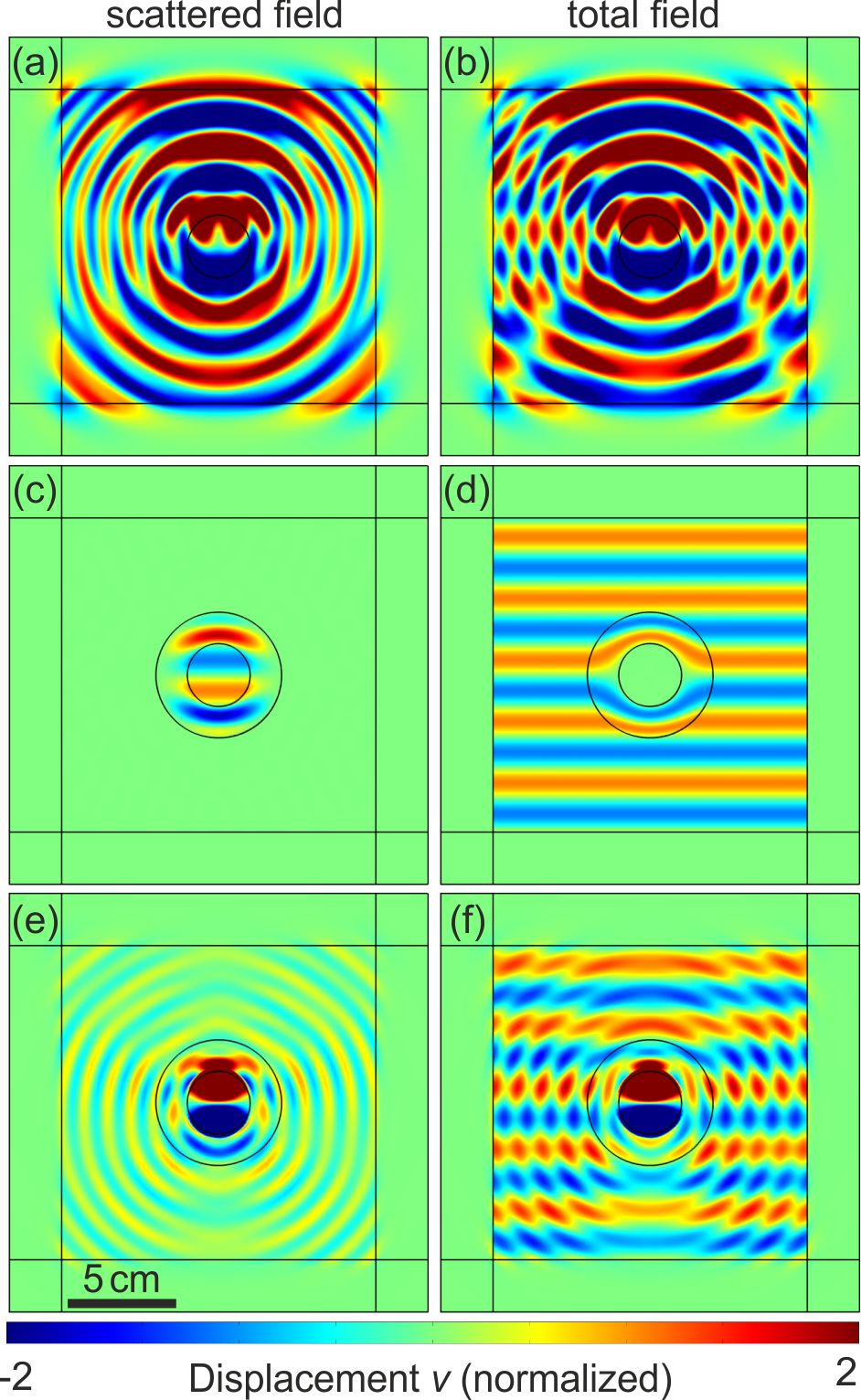}
\caption{Scattered (left) and total (right) fields for diffraction by (a), (b) an obstacle  as in Fig.\,2(c),  ~ (c), (d) an ideal Cosserat cloak and  (e), (f) an obstacle dressed with symmetrized cloak, under a compressional excitation at $2\times 10^5$ rad.$\rm s^{-1}$. Precisely, the real part of the vertical component of the displacement vector, $v$, normalized to the amplitude of the incident wave, is shown on a false-color scale. All cloaks have
an inner radius $r_1=0.03$ m and outer radius $r_2=0.06$ m.
}
\label{fig3}
\end{figure}


\section{Conclusion}
In conclusion, we have introduced general perfectly matched layers and a scattered-field formalism for elastodynamic waves following the Navier equations. We have shown that the adaptative perfectly matched layers work all the way from the quasi-static regime to high frequencies. Different types of boundary conditions have been discussed. This mathematical progress should be useful for many different mechanical problems. As an example, it has enabled us to analyze quantitatively the scattering reduction of elastodynamic cloaks. Our numerical results indicate ideal cloaking for Cosserat-type elasticity tensors that do not obey the minor symmetries -- which is expected from the analytical construction -- whereas approximative continuous symmetrized elasticity-tensor distributions lead to a rather poor cloaking quality. This shortcoming is connected to the fact that symmetric elasticity tensors cannot simultaneously deal with incident compression and shear waves whenever the density is isotropic.
 It would thus be highly desirable to construct and realize Cosserat mechanical metamaterials experimentally in the future.

\section*{References}

\end{document}